\documentstyle[epsf]{article}

\def\abstract#1{\vskip 7mm 
        \begin{center}{\large Abstract}\par \smallskip
                \begin{minipage}[c]{12cm}
                        \small #1
                \end{minipage}
        \end{center}
}
\def\title#1{\begin{center}{\Large\bf #1}\end{center}}
\def\author#1{\vskip 5mm \begin{center}{#1}\end{center}}
\def\address#1{\begin{center}{\it #1}\end{center}}
\makeatletter
\@ifundefined{lesssim}{}{}
\@ifundefined{gtrsim}{}{}
\def\vereq#1#2{\lower3pt\vbox{\baselineskip1.5pt \lineskip1.5pt
\ialign{$\m@th#1\hfill##\hfil$\crcr#2\crcr\sim\crcr}}}
\makeatother

\begin{document}

\title{%
   An Open Universe from Valley Bounce
  \smallskip \\
}
\author{%
  Kazuya Koyama\footnote{E-mail:kazuya@phys.h.kyoto-u.ac.jp}
}
\address{%
 Graduate School of Human and Environment Studies, \\ 
 Kyoto University, Kyoto  606-8501, Japan
}
\author{%
 Kayoko Maeda\footnote{E-mail:maeda@phys.h.kyoto-u.ac.jp}
 Jiro Soda\footnote{E-mail:jiro@phys.h.kyoto-u.ac.jp}
}
\address{
 Department of Fundamental Sciences, FIHS, Kyoto University,\\
       Kyoto, 606-8501, Japan
}
\abstract{
  It appears difficult to construct a simple model
for an open universe based on the one bubble inflationary scenario.
The reason is that one needs a large mass to avoid the
tunneling via the Hawking Moss solution and a small mass for successful 
slow-rolling. However, Rubakov and  Sibiryakov suggest that the 
Hawking Moss solution is not a solution for the false vacuum decay 
process since it does not satisfy the boundary condition. 
Hence, we have reconsidered the arguments for the defect of 
the simple polynomial model. We find the valley bounce belonging
to a valley line in the functional space represents the decay process
instead of the Hawking Moss solution. 
The point is that the valley bounce gives the appropriate
initial condition for the inflation.
We show an open inflation model 
can be constructed within the polynomial form of the potential so
that the fluctuations can be reconciled with the observations.
Details of the analysis can be seen in Ref.\cite{KK}. }


\section{Introduction}
\hspace{1cm}

Recent observations suggest the matter density of the universe is 
less than the critical density. 
Hence, it is desirable to have a model for an open universe, 
say $\Omega_0 \sim 0.3$. The realization of an open universe is
difficult in the ordinary inflationary scenario. This is because
if the universe expands enough to solve the horizon problem, 
the universe becomes almost flat. One attempt to realize 
an open universe in the inflationary scenario is to consider inside 
the bubble created by the false vacuum decay \cite{Gott}. 
The scenario is as follows. Consider the
potential which has two minimum. One is the false vacuum which has 
non-zero energy and the other is the true vacuum. Initially the field is 
trapped at the false vacuum. Due to the potential energy, universe expands
exponentially and the large fraction of the universe becomes
homogeneous. As the false vacuum is unstable, it decays and creates 
the bubble of the true vacuum. If the decay process is well suppressed, 
the interior of the bubble is still homogeneous. 
The decay is described by the $O(4)$ symmetric configuration 
in the Euclidean spacetime.
Then, analytical continuation of this configuration 
to the Lorentzian spacetime 
describes the evolution of the bubble which looks from the inside like
an open universe.
Unfortunately, since the bubble radius cannot be greater than the
Hubble radius, the created universe is curvature dominated even if the
whole energy of the false vacuum is converted to the energy of the matter
inside the bubble \cite{Yokoyama1}. 
Thus, the second inflation in the bubble is needed. 
If this second inflation stopped when $\Omega<1$, our universe becomes 
homogeneous open universe. 

Though the basic idea is simple, the realization of this scenario
in a simple model has been recognized difficult \cite{Linde1}.
The difficulty is usually explained as follows.
Consider the model involving one scalar field. 
For the polynomial form of the potential like 
$V(\phi)= m^2\phi^2- \delta \phi^3 + \lambda \phi^4$, 
the tunneling should occur at sufficiently large $\phi$ 
to ensure that the second inflation gives the appropriate density parameter. 
Then, the curvature around the barrier which separates the false and
the true vacuum is small compared with the Hubble
scale which is determined by the energy of the false vacuum. In this case,
the field jumps up onto the top of the barrier due to the quantum diffusion.  
When the field begins to roll down from the top of the 
barrier, large fluctuations are formed due to the quantum diffusion at
the top of the barrier. Then the whole scenario fails. This problem is
rather generic. To avoid jumping up, the curvature around the barrier
should be large compared with the Hubble scale $V''> H^2$. On the other
hand, to realize the second inflation, the field should roll down 
slowly, then we need $V''<H^2$. These two conditions are incompatible.

There are several attempts to overcome this problem. 
Recently Linde constructs the potential which has sharp peak near the 
false vacuum \cite{Linde2}. In this potential, 
the tunneling occurs and at the same time slow-rolling is allowed after 
tunneling, then the second inflation can be realized. 
But, it is still unclear what is the physical mechanism for the appearance 
of the sharp peak in the potential. 

We will reconsider this problem from different perspective. 
The point is the understanding of the tunneling process. 
In the imaginary-time path-integral formalism, tunneling is 
described by the solution of the Euclidean field equation. 
This solution gives the saddle-point of the path-integral. 
Then this determines the semi-classical exponent of the
decay rate $\exp(-S_E(\phi_B))$, where $S_E$ is the Euclidean action. 
In the case the curvature around the barrier is small compared with the
Hubble, the solution is given by the Hawking Moss (HM) solution, 
which stays at the top of the barrier through the whole Euclidean time 
\cite{HM}. 
Recently Rubakov and  Sibiryakov give the interpretation
of this tunneling mode using the constrained instanton method 
\cite{Rubakov,Affleck}. They show the HM solution does not represent
the false vacuum decay if one takes into account the analytic continuation to
the Lorentzian spacetime. This is because this solution does not satisfy the
boundary condition that the field exists in the false vacuum at the
infinite past. However, this does not imply the decay does not occur.
One should consider a family of the almost saddle-point 
configurations instead of the true solution of the Euclidean
field equation.  They show although the decay rate is determined by the HM
solution, the structure of the field after tunneling is determined by
the other configuration which is one of the almost saddle-point solutions.
In this method, one must choose the constraint so that a family of
almost saddle-point solutions well covers the region which is expected to 
dominate the path-integral. One way to realize this is to cover the
valley region of the functional space of the action 
\cite{Balitsky, Aoyama1}.  
Along the valley line, the action varies most gently. 
Then it is reasonable to take the
configurations on the valley line as a family of the almost saddle-point 
configurations. We will call the configuration on the valley line
of the action the valley bounce $\phi_V$.

This analysis gives the possibility to overcome the problem. Even if 
the curvature around the barrier is small compared with the Hubble
scale, it implies there is a possibility to occur the tunneling
described by the valley bounce. 
If the field appears sufficiently far from the top, one can avoid 
the large fluctuations. During the tunneling, fluctuations of the tunneling
field are generated. These fluctuations are stretched during the second
inflation and observed in the open universe. These should be compatible
with the observation. Once this can be confirmed, there is no difficulty
in constructing the one bubble open inflationary model in the simple model
with the polynomial form of the potential. 

In this paper, we show this is true as long as the tunneling is described
by the valley bounce. We clarify the structure of the valley 
bounce extending the method developed by Aoyama.et.al \cite{Aoyama1} 
to the de Sitter spacetime. 
We show the fluctuations can be reconciled with the observations.


\section{Valley method in de Sitter spacetime}
\hspace{1cm}

First we review the formalisms which are necessary to describe the 
false vacuum decay in the de Sitter space.
We want to examine the case 
in which the gravity comes to play a role. Unfortunately, 
we have not known how to deal with quantum gravity effect yet. 
So, we study the case in which we can treat gravity at the semi-classical 
level. That is, we treat the problem within the framework of the field 
theory in a fixed curved spacetime \cite{Rubakov}. 
The potential relevant to the tunneling is given by
\begin{equation}
V(\phi)=\epsilon+V_T(\phi).
\end{equation}
We assume $\epsilon$ is of the order 
$M_{\ast}^4$ and $V_T(\phi)$ is of the order $M^4$. 
We study the case $M$ is small compared to $M_{\ast}$,  $M \ll  M_{\ast}$.
Then the geometry of the spacetime is fixed to the de Sitter spacetime
with $H=M_{\ast}^2/M_p$, where $M_p^{-2}=8 \pi G/3$.
We consider the situation in which the potential $V_T(\phi)$ has the false 
vacuum at $\phi=\varphi_F$ and the top of the barrier at $\phi=\varphi_T$. 
Since the background metric is fixed, we can change 
the origin of the energy freely. We choose $V_T(\phi_F)=0$. 
Following, we work in units with $H=1$.

The decay rate is given by the imaginary part of the path-integral
\begin{equation}
Z=\int [d \phi] \exp {\left(-S_E(\phi) \right)},
\end{equation}
where $S_E$ is the Euclidean action relevant to the tunneling.
The dominant contribution of this path-integral is given by 
the configurations which have $O(4)$ symmetry \cite{CL}. So, we assume the 
background metric and the field to have the form
\begin{eqnarray}
ds^2 &=& d \sigma^2+a(\sigma)^2 \left( d\rho^2+\sin^2 \rho
d \Omega \right), \nonumber\\
\phi &=& \phi(\sigma),
\end{eqnarray}
where $a(\sigma)=\sin \sigma$. 
Then, the Euclidean action of $\phi(\sigma)$ is given by
\begin{equation}
S_E = 2 \pi^2 \int d \sigma    \left( a^3
\left( \frac{1}{2}\phi'^2+V_T(\phi) \right) \right).
\end{equation}  

The saddle-point of this path-integral is determined by 
the Euclidean field equation $\delta S_E/\delta \phi$=0;
\begin{equation}
\phi''+3 \cot \sigma \: \phi'-V_T'(\phi)=0.
\end{equation}
We impose the regularity conditions at the time when $a(\sigma)=0$ as
\begin{equation}
\phi'(\sigma=0)=\phi'(\sigma=\pi)=0.
\end{equation}
We represent the solution of this equation as $\phi_B(\sigma)$.
If the fluctuations around this solution have a negative mode, this
gives the imaginary part to the path-integral and this solution
contributes to the decay dominantly. The decay rate $\Gamma$ is evaluated by
\begin{equation}
\Gamma \sim \exp (-S_E(\phi_B)).
\end{equation}
The equation has two types of the solutions depending on the shape of 
the potential.
If the curvature around the barrier is large compared with the Hubble
scale, then the Coleman De Luccia (CD) solution  
and the Hawking Moss (HM) solution exist \cite{HM,CL}. 
In this case, the decay is described by the CD solution. The analytic
continuation of this solution to Lorentzian spacetime describes the
bubble of the true vacuum. On the other hand, in the case the curvature 
around the barrier is small compared with the Hubble scale, 
only the HM solution exists. This solution is a trivial solution 
$\phi=\varphi_T$. The meaning of the HM solution is somewhat ambiguous. 
There are several attempts to interpret this tunneling mode. 
One way is to use the stochastic approach \cite{Linde3}. 
Within this approach, 
it has been demonstrated that the decay rate given by eq.(7) coincides 
with the probability of jumping from the false vacuum $\varphi_F$ onto the
top of the barrier $\varphi_T$ due to the quantum fluctuations. 

Recently, Rubakov and Sibiryakov give the interpretation of the 
HM solution using the constrained instanton method
\cite{Rubakov,Affleck}. The main idea is to
consider a family of the almost saddle-point configurations instead
of the true solution of the Euclidean field equation, i.e. the HM solution. 
The motivation comes from the boundary condition. They take the boundary 
condition that the state of the quantum fluctuations above the classical 
false vacuum is the conformal vacuum. In this case they show 
the field should not be constant at $0 < \sigma < \pi$ and the HM
solution is excluded by this boundary condition. 
Then one should seek the other configurations which obey the boundary
condition and dominantly contribute to the path-integral.
In the functional integral, the saddle-point solution gives the most
dominant contribution, but the contribution from a family of almost
saddle-point configurations which have almost the same action with
that of the saddle-point solution should also be included. 
To realize this in the functional integral, one introduces the 
identity $1=\int d \alpha \delta({\cal C}-\alpha)$ into the path
integral for some constraint ${\cal C}$. First choose one $\alpha$.
This selects the subspace of the functional space. In this subspace,
we can perform the integral of the field using saddle-point method under 
the constraint. The minimum in this subspace satisfies the equation of motion 
with constraint instead of the field equation. This minimum corresponds
to the almost saddle-point configuration $\phi_{\alpha}$ 
which is slightly deformed from the HM solution. Changing $\alpha$, these
configurations form a trajectory. We can evaluate the path-integral 
by integrating over $\alpha$ along this trajectory. Since along this
trajectory the HM solution gives the minimum action, integrating over 
$\alpha$ gives the decay rate determined by the HM solution. But the
structure of the field after tunneling can be determined 
by the other configuration on this trajectory $\phi_{\alpha}$.
They found the configuration which describes the bubble of the true vacuum 
if we continue it to the Lorentzian spacetime. 
Then, they conclude that even 
in the case only the HM solution exists, the result of the tunneling 
process can be the bubble of the true vacuum which is described by one of
the almost saddle-point configurations.

In this formalism, the validity of the method depends on the choice of
the constraint \cite{Aoyama1,Aoyama2}. 
This is because, in practice, we do the Gaussian integral
around the almost saddle-point solutions. To evaluate the path-integral 
properly we should choose the constraint so that a family of
almost saddle-point solutions well covers the region which is expected to 
dominate the path-integral. Since the action varies most gently along
the valley line, one way to realize the aim 
is to cover the valley region of the action \cite{Balitsky, Aoyama1}.  
One can identify the configurations on the valley line and make
Gaussian integral around these configurations. 
 
Taking into account the above fact, it is desirable to analyze the 
structure of not only the solution of the Euclidean field equation
but also the configurations on the valley line.
One way to define the configurations on this valley line is
to use the valley method developed by Aoyama.et.al \cite{Aoyama1}.
To obtain the intuitive understanding of this method, consider the
system of the field $\phi_i$. Here $i$ stands for the discretized
coordinate label and we take the metric as $\delta_{ij}$. 
In the valley method the equation which identifies the valley
line in the functional space is given by
\begin{equation}
D_{ij} \partial_i S = \lambda \: \partial_i S,\:\:\:\: D_{ij}=\partial_i 
\partial_j S,
\end{equation}
where $\partial_i=\partial/\partial \phi_i$. Since this equation has 
one parameter $\lambda$, this defines a trajectory in the space of
$\phi$. The parameter $\lambda$ is one of the eigen value of the matrix
$D_{ij}$. On this trajectory the gradient vector $\partial_i S$ is
orthogonal to all the eigenvectors of $D_{ij}$ except for the
eigenvector of the eigen value $\lambda$.
This equation can be rewritten as
\begin{equation}
\partial_i \left( \frac{1}{2}(\partial_j S)^2-\lambda S \right)=0.
\end{equation}
This allows the interpretation of the solution for the equation. It
extremizes the norm of the gradient vector $\partial_i S$ 
under the constraint $S=$const., where
$\lambda$ plays the role of the Lagrange multiplier. Such solution
can be found each hypersurface of constant action, then the solutions
of the equation form a line in the functional space. 
If we take $\lambda$ as the 
one with the smallest value, then the gradient vector is minimized.
In this case, the action varies most gently along this line. 
This is a plausible definition of the valley line. 
We will call the configuration on the valley line of the
action the valley bounce $\phi_V$ and the trajectory they form the valley
trajectory. 

Following we formulate this method in the de Sitter spacetime.
The most convenient way is to use the variational method eq.(9).
We shall define the valley action by
\begin{equation}
S_V = S_E-\frac{1}{2 \lambda} \int d \sigma  \sqrt{g} 
\left( \frac{1}{\sqrt{g}} \frac{\delta S_E}{\delta \phi} \right)^2.
\end{equation}
The valley bounce is obtained by varying this action.
The equation which determines the valley bounce $\delta S_V/\delta \phi=0$
is a fourth order differential equation. 
We introduce the auxiliary field $f$ to cancel the fourth derivative
term \cite{Aoyama3};
\begin{equation}
S_{f}=\frac{1}{2 \lambda} \int d \sigma \sqrt{g}
\left( f-\frac{1}{ \sqrt{g}} \frac{\delta S_E}{\delta \phi} \right)^2.
\end{equation}
Then the valley action becomes
\begin{equation}
S_V+S_f = S_E+\frac{1}{2 \lambda} \int d \sigma \sqrt{g} f^2 
-\frac{1}{\lambda} 
\int d \sigma f \frac{\delta S_{E}}{\delta \phi}.
\end{equation}
Taking the variation of this action with respect to $f$ and $\phi$, we obtain
the equations for $\phi$ and $f$;
\begin{eqnarray}
\frac{1}{\sqrt{g}} \frac{\delta S_E}{\delta \phi} &=& f,
\nonumber\\
\int d \sigma'
\frac{\delta^2 S_E}{\delta \phi(\sigma) \delta \phi(\sigma')} f(\sigma)
&=& \lambda  \sqrt{g} f(\sigma).
\end{eqnarray}
Using $a(\sigma)=\sin \sigma$, the valley equation 
which determines the structure of the valley bounce is given by
\begin{eqnarray}
\phi''+3 \cot  \sigma \: \phi'- V_T'(\phi) &=& -f,  \nonumber\\
f''+3 \cot  \sigma  \: f'-V_T''(\phi)f &=& - \lambda f.
\end{eqnarray}

We analyze the structure of the valley bounce for the case only the
HM solution exists. We construct the piece-wise quadratic potential
in which we can solve the valley equations analytically.
The potential which we study is
\begin{eqnarray}
 V_T(\phi) = \left\{ 
\begin{array}{ll}
\frac{1}{2}m_F^2(\phi-\varphi_F)^2,&
\qquad -\infty <\phi < 0,  \\
\\
-\frac{1}{2}m_T^2(\phi-\varphi_T)^2+\eta,&
\qquad 0 \leq \phi< \infty, \end{array}
\right.
\end{eqnarray}
where $\eta$ is of the order $M^4$.
For $m_T^2<4$, only the HM solution exists. For example we take $m_T^2=2$, 
$m_F^2=0.5$ and $\eta=0.1 M^4$. The HM solution has one negative
eigenvalue $\rho_{HM,-}=-2$ and the smallest positive eigenvalue is given by
$\rho_{HM,+}=2$. The generic feature of the valley bounce is
understood by the simple analysis of the case in which the valley bounce
exists only in one parabola. First consider the valley trajectory associated 
with the negative eigenvalue. The solution of the valley equation is 
essentially has a form $f=\lambda (\phi-\varphi_T)=\mbox{const}$. 
This solution does not represent the tunneling, so we seek the trajectory
associated with the smallest positive eigenvalue 
$\lambda(\phi_{HM})=\rho_{HM,+}$. 
The solution of the valley equation is given by
$\phi-\phi_T \propto \cos \sigma$ and $f = \lambda (\phi-\phi_T)$ (Fig.1).
In this trajectory, the HM solution gives the minimum of the action 
(Fig.2). The horizontal coordinate is the norm of the filed 
$\Phi=\sqrt{\int d \sigma a(\sigma)^3 \vert \phi(\sigma)-\varphi_T \vert^2}$.
The action grows as the variation of the
field becomes large, but this increase is relatively gentle.

Although the HM solution gives the dominant contribution to the
path-integral, this solution does not satisfy the boundary condition for
the false vacuum decay as shown by Rubakov and Sibiryakov \cite{Rubakov}.
Making Analytic continuation to the Lorentzian spacetime at $\sigma=0
(z=-1)$, the field moves according to the field equation. 
If the field reaches $\varphi_F$, this solution represents
the false vacuum decay. The behavior of the field in this Lorentzian
spacetime is determined by the initial position of the
field. This is determined by the behavior of the field at $\sigma=0$ 
in the Euclidean region. Provided that its initial position is different
from $\varphi_T$, this boundary condition can be satisfied. From this
fact, the HM solution does not satisfy the boundary condition. On the
other hand the valley bounce does satisfy the boundary condition.
Furthermore the fluctuations around the valley bounce should have 
one negative mode to ensure that the valley bounce plays a role instead
of the HM solution. The valley bounce has a lowest eigenvalue 
$\rho_{V,-} < \lambda(\phi_V)$. We find this is negative on this trajectory.
Since this is the unique negative eigenvalue, the gaussian integration of
the fluctuations around this valley bounce gives the imaginary part to
the path-integral. Then, the valley bounce contributes 
to the false vacuum decay and describes the creation of the bubble of
the true vacuum.

\section{An open universe from valley bounce}
\hspace{0.1cm}

We will see an open inflation model can be constructed using the valley 
bounce. Following, we restore the Hubble scale $H$. 
Since the radius of the bubble $R$ is small compared with the Hubble
horizon \cite{CL}, then the curvature scale is greater than the energy 
of the matter inside the bubble $\rho_M$ even if the whole energy
of the false vacuum is converted to it, $\rho_M/M_p^2 \sim H^2 <1/R^2$
\cite{Yokoyama1}.
Then, we need the second inflation in the bubble. To realize the second
inflation inside the bubble, the field should roll slowly down the
potential. This implies the curvature of the potential is small 
compared with the Hubble. To avoid the $ad$ $hoc$ fine-tuning of the 
potential, we will assume this is true for all region of the potential.
In this case, since $m_T <H $ the solution of the Euclidean equation
is given by the HM solution and the valley bounce is shown as in Fig.7.
We connect the linear potential at the point the field appears after the
tunneling $\phi=\phi_{\ast}$,
\begin{equation}
V(\phi)=V_{\ast}- \mu^3 (\phi-\phi_{\ast}), \qquad (\phi> \phi_{\ast}).
\end{equation}
We demand the potential and its derivative are connected smoothly
at the connection point $\phi_{\ast}$. Then we obtain
\begin{eqnarray}
V_{\ast} &=&\epsilon+\eta-\frac{1}{2}m_T^2(\phi_{\ast}-\varphi_T)^2, 
\nonumber\\
\mu^3 &=& m_T^2 (\phi_{\ast}-\varphi_T).
\end{eqnarray}
The initial conditions of the field are given by the valley bounce
\begin{equation}
\phi(t=0)=\phi_0(z=1)=\phi_{\ast},\:\:\: \dot{\phi}(t=0)=0.
\end{equation}
If the field obeys the classical field equation;
\begin{equation}
\ddot{\phi}+3 \coth t \: \dot{\phi}+V'(\phi)=0,
\end{equation}
then the solution of $\phi$ satisfies 
\begin{equation}
\dot{\phi}(t)=\mu^3 \: \frac{\cosh^3 t-3 \cosh t +2}{3 \sinh^3 t}.
\end{equation}
In the small $t$ this behaves as $(1/4) \mu^3 t$.
The classical motion during one expansion time is given by
$\vert \dot{\phi} \vert H^{-1}$. 
On the other hand the amplitude of the
quantum fluctuations is given by $\delta \phi \sim H$. The curvature
perturbation ${\cal R}$ produced by the quantum fluctuations is approximately
given by the ratio of these two quantities; 
\begin{equation}
{\cal R} \sim \frac{\delta \phi}{\vert \dot{\phi} \vert H^{-1}}
\sim \frac{H^3}{\mu^3} \sim \frac{H^2}{m_T^2} \left(\frac{H}{\phi_\ast
-\varphi_T} \right).
\end{equation}
This should be of the order $10^{-5}$ from the observation of the 
cosmic microwave background (CMB) anisotropies.
If $\vert \phi_{\ast}-\varphi_T \vert < H$, as in the case the HM
solution describes the tunneling, ${\cal R} >1$ and the scenario cannot
work well. This is because at $\phi_{\ast} \sim \varphi_T$, 
the field experiences the quantum diffusion rather than 
the classical potential force.
Fluctuations in this diffusion
dominated epoch make the inhomogeneous delay of the start of the
classical motion, thus make large fluctuations. Fortunately, from Fig.7,
we see for appropriate $\lambda$, the valley bounce gives the initial 
condition as $\vert \phi_{\ast}-\varphi_T \vert \sim O(1)(M^2/m_T)$, 
which is larger than the Hubble if $M >H$. In this case, the potential force
works and the field rolls slowly down the potential. 
We expect the curvature perturbation can be suppressed 
for the valley bounce. In fact, we find the power of the 
curvature purturbations is given by
\begin{equation}
\lim_{p \to \infty} \frac{p^3}{2 \pi^2} 
P_{{\cal R}}(p,\lambda) = \frac{1}{4 \pi^2} \left(\frac{3 H^3}{\mu^3}
\right)^2 \sim 
\left(\frac{M_{\ast}^2}{M_p M} \right)^4 \left(\frac{H}{m_T} \right)^2.
\end{equation}
Here we use the fact the valley bounce gives the initial condition as
$\vert \phi_{\ast}-\varphi_T \vert \sim M^2/m_T$, then 
$\mu^3 =m_T M^2$. This quantity should be of the order $10^{-10}$ 
from the observation. This can be achieved by taking 
$(M^2_{\ast}/M) \ll M_p$.

\section{Conclusion}
\hspace{0.1cm}

It is difficult to provide the model which solves the horizon problem
and at the same time leads to the open universe in the context of the
usual inflationary scenario. In the one bubble open inflationary
scenario, the horizon problem is solved by the first inflation and
the second inflation creates the universe with the appropriate $\Omega_0$.

Many works have been done within this framework of the 
scenario and it is recognized this scenario requires additional
fine-tuning \cite{Linde1,Linde2}. 
The defect is thought to arise because the curvature 
around the barrier should be larger than the Hubble scale 
to avoid large fluctuations, which contradicts to the requirement that 
the curvature of the potential should be small to realize the second 
inflation inside the bubble. 

Thus to complete the scenario, we should solve this problem.
The main claim of this paper is that this problem can be solved
in the simple model with the polynomial form of the potential.
We reconsidered the tunneling process from the different perspective.
If the curvature around the potential is small, the tunneling is
described by one of a family of the almost saddle-point solutions
\cite{Rubakov}. 
This is because the true saddle-point solution, that is, the Hawking Moss
solution does not satisfy the boundary condition for the false vacuum decay.
The main idea is that the almost saddle-point solution can 
give the appropriate initial condition for the second inflation. 
A family of the almost saddle-point solutions generally forms 
a valley line in the functional space. We called the configurations on
the valley line valley bounces. 
To identify valley bounces, we applied the valley method 
developed by Aoyama.et.al \cite{Aoyama1}. 
In this method these configurations can be 
identified using the fact the trajectory they form in the functional
space corresponds to the line on which the action varies
most gently. We formulated this method in the de Sitter spacetime and 
clarified the structure of the valley bounces. We found the
valley bounce which gives the appropriate initial condition of the
second inflation even if the curvature around 
the barrier is small compared with the Hubble scale.
Consider the case this valley bounce describes the tunneling.
It is possible the field appears sufficiently far from the top
of the barrier after the tunneling, 
then we can avoid the large fluctuations. 
Hence, using the valley bounce, we can solve the
problem which arises in the open inflationary scenario besides the usual 
fine-tuning of the inflationary scenario.  
The one bubble open inflation model can be constructed without difficulty.

\section*{Acknowledgements}
The work of J.S. was supported by Monbusho Grant-in-Aid No.10740118
and the work of K.K. was supported by JSPS Research Fellowships for
Young Scientist No.04687


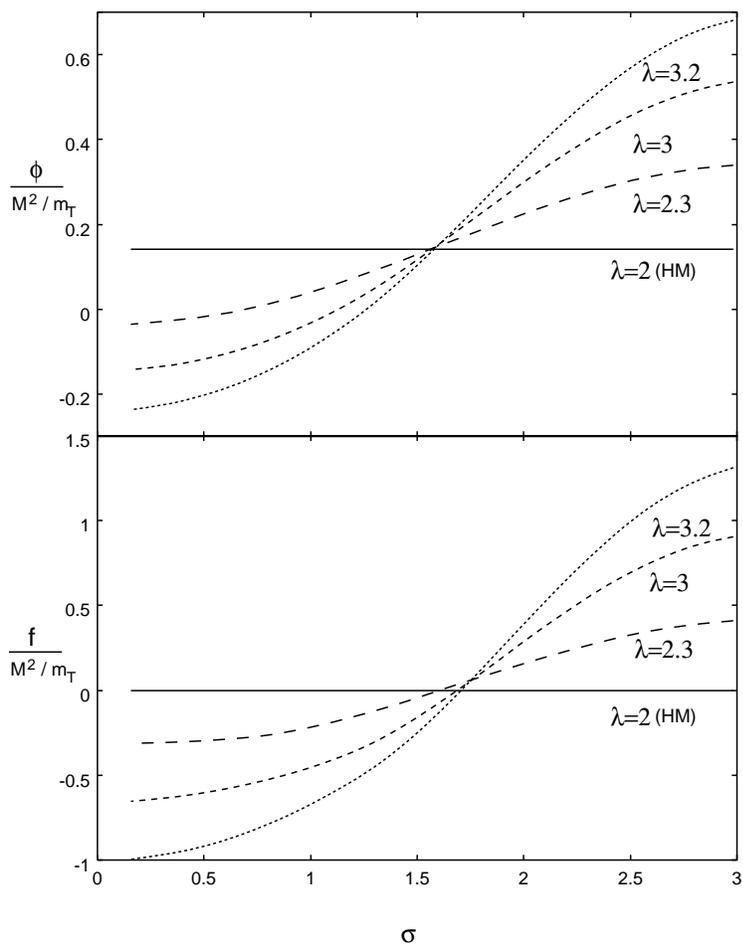
\begin{figure}[h]
 \epsfysize=12.5cm  
 \begin{center}
  \epsfbox{fig7.eps }
 \end{center}
 \caption{}
 \end{figure}

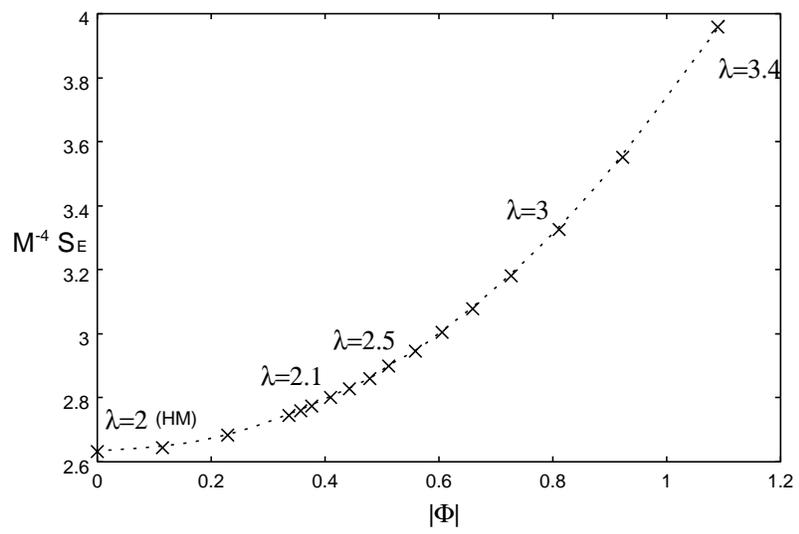
\begin{figure}[h]
 \epsfysize=7cm  
 \begin{center}
  \epsfbox{fig6.eps }
 \end{center}
 \caption{}
 \end{figure}


\begin{thebibliography}{99}
\bibitem{KK} K. Koyama, K. Maeda and J. Soda, hep-ph/9910556.

\bibitem{Gott} J.R. Gott III,
Nature {\bf 295},   304  (1982);\\ J.R. Gott III and T.S. Statler,
Phys. Lett. {\bf B136},  157   (1984).
%
\bibitem{Yokoyama1} M. Sasaki, T. Tanaka, K. Yamamoto, and J. Yokoyama,  
Phys. Lett. {\bf B317}, 510 (1993).
%
\bibitem{Linde1} A.D. Linde,     Phys. Lett.   {\bf B351}, 99 (1995);
A.D. Linde and A. Mezhlumian, Phys. Rev. D {\bf 52}, 6789 (1995).
%
\bibitem{Linde2} A.D. Linde,     Phys. Rev.   {\bf D59}, 023503 (1999);
%
\bibitem{HM} S.W. Hawking and I.G. Moss, Phys. Lett.  {\bf B110},  35
(1982).
%
\bibitem{Rubakov} V.A. Rubakov and S.M. Sibiryakov,
	preprint, gr-qc/9905093 (1999).
%
\bibitem{Affleck} I. Affleck, Nucl. Phys. {\bf B191}, 429 (1981). 
%
\bibitem{Balitsky} I.I. Balitsky and A.V. Yung,
Phys. Lett.   {\bf B168}, 113 (1986).
\bibitem{Aoyama1}
H. Aoyama and H. Kikuchi, Nucl. Phys. {\bf B369}, 
219 (1992);
for a review see H. Aoyama, T. Harano, H. Kikuchi, I. Okouchi, M. Sato 
and S. Wada, Prog. Theor. Phys. Suppl. {\bf 127}, 1 (1997).
%
\bibitem{CL} S. Coleman and F. De Luccia,   Phys. Rev. {\bf D21},  3305
(1980).
%
\bibitem{Linde3} A. Goncharov and A. Linde,  Sov. J. Part. Nucl.
{\bf 17}, 369 (1986);  
A.D. Linde,   Nucl. Phys. {\bf B216}, 421  (1983).
%
\bibitem{Aoyama2} H. Aoyama, T. Harano, M. Sato and S. Wada, 
Nucl. Phys. {\bf B466}, 127 (1996).
%
\bibitem{Aoyama3} H. Aoyama and S. Wada, Phys. Lett. {\bf B349}, 279 (1995).
%

\end{thebibliography}
\end{document}